\documentclass{osa-article}

\usepackage{units,upgreek}
\usepackage{hyperref}

\usepackage[caption=false]{subfig}
\journal{oe}
\usepackage{amsmath}
\usepackage{amssymb}
\usepackage{cleveref}
\let\oldref\ref
\renewcommand{\ref}[1]{(\oldref{#1})}

\begin{document}
\newcommand{\h}[1]{\hat{#1}}
\renewcommand{\d}{\mathrm{d}}
\newcommand{\sinc}{\mathrm{sinc}}

\title{A high dynamic range optical detector for measuring single photons and bright light}

\author{Johannes Tiedau,\authormark{*} Evan Meyer-Scott, Thomas Nitsche, Sonja Barkhofen, Tim J. Bartley, and Christine Silberhorn}

\address{Applied Physics, University of Paderborn, Warburger Stra\ss e 100, 33098 Paderborn, Germany}

\email{\authormark{*}johannes.tiedau@upb.de} 



\begin{abstract}
Detecting light is fundamental to all optical experiments and applications. At the single photon level, the quantized nature of light requires specialised detectors, which typically saturate when more than one photon is incident.
Here, we report on a massively-multiplexed single-photon detector, which exploits the saturation regime of a single click detector to exhibit a dynamic range of \unit[123]{dB}, enabling measurement from optical energies as low as $10^{-7}$ photons per pulse to $\sim2.5\times10^{5}$ photons per pulse. This allows us to calibrate a single photon detector directly to a power meter, as well as characterize the nonclassical features of a variety of quantum states. 
\end{abstract}

\section{Introduction}

Optical detectors are based on a broad range of physical principles, which dictate the range of powers to which they are sensitive. The dynamic range of an optical detector is defined as the difference between its noise floor and saturation intensity. Above the saturation intensity, the detector response is constant, such that different light levels cannot be distinguished. We differentiate saturation from the breakdown intensity of the detector, namely pulse energies above which the detector response is permanently changed ({i.e.} latched or damaged). For different optical detectors, the saturation intensity and breakdown intensity is determined by its principle of operation. 
For example, ideal single photon binary (``click, no-click'') detectors are saturated when at least one photon is incident. As such, single-photon level detectors cannot be used to measure pulse energies beyond one photon. To overcome this limitation, multiplexing schemes are used to divide an incoming pulse such that the average intensity per multiplexing bin is below the saturation level. Existing multiplexed single-photon detection schemes~\cite{paul_photon_1996,kim_multiphoton_1999,
banaszek_photon_2003,achilles_fiber-assisted_2003,rehacek_multiple-photon_2003,fitch_photon-number_2003,achilles_photon-number-resolving_2004,
castelletto_reduced_2007,schettini_implementing_2007,the_multiple-photon_2007,divochiy_superconducting_2008,micuda_high-efficiency_2008,brida_scalable_2009,pomarico_room_2010,
allman_near-infrared_2015} still suffer from saturation effects in that they are not sensitive to photon numbers greater than the number of bins in the multiplexed device~\cite{kruse_limits_2017}. Photon-number-resolving detectors also suffer from saturation, however this may be overcome by careful analysis of the detector response function~\cite{gerrits_extending_2012}. 

Here, we show that we can overcome this saturation limitation, and extend the sensitivity of single-photon binary detectors to 250000 photons per pulse. To do this, we use a detector coupled to a loop of fibre, as first introduced by Banaszek and Walmsley~\cite{banaszek_photon_2003}. Below the saturation limit, this detector architecture generates a logarithmic response to the incoming pulse energy~\cite{the_multiple-photon_2007,webb_photostatistics_2009}. However, we show that this device can be pushed beyond its saturation level, and information on the incident pulse energy can still be extracted, limited only by the breakdown level of the click detector. This produces a massively-multiplexed detector with a dynamic range of \unit[123]{dB}. We perform measurements at low photon-numbers, where this detector is capable of differentiating different quantum states based on their photon statistics, and at high pulse energy, whereby the same device can be directly compared with a power meter. This provides a traceable comparison at the single-photon level to optical standards without requiring calibrated attenuation~\cite{marsili_detecting_2013} or synchrotron radiation~\cite{muller_traceable_2014,mueller_verification_2017}.

\begin{figure}[!h]
\centerline{\includegraphics[width=0.7\linewidth]{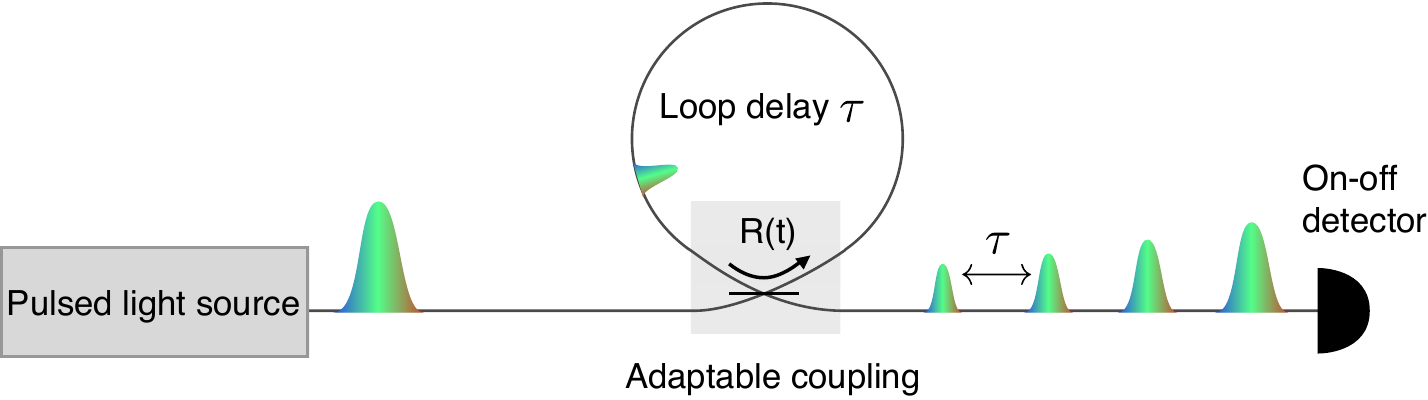}}
\caption{Binary detector coupled to a resonator with coupling $R(t)$. For active switching, $R(t=0)=0$, {i.e.} all the light is switched into the loop. Subsequently, $R(t)=R$ is constant. In the passive case, $R(t)=R$ is constant for all times.}
\label{fig.overview}
\end{figure}

\section{Time-multiplexed detector design}
Our multiplexed detector 
comprises a variable beam splitter with one output connected to one input, and the other output connected to a single-photon detector, as shown in Fig.~\ref{fig.overview}. The variable beam splitter can be active, where the entire pulse is switched into the loop, before decaying, or passive, whereby the reflectivity of the beam splitter is constant and part of the initial pulse energy is reflected directly to the detector, comprising the first bin. 
For an input pulse of given intensity, an output pulse train with a characteristic decay of optical energy is produced. 
We ensure the pulse duration and dead time of the detector are much shorter than the round trip time of the loop, discretizing the ring-down in a series of time bins $j$. 

The amount of light coupled to the first bin depends on whether active or passive switching of the light pulse into and out of the loop is used.
In each subsequent bin some fraction of light remaining in the loop is coupled to the detector. 
If this energy is below the saturation level of the detector, the probability of a click in a given bin $j$ follows an exponential decay, as previously studied~\cite{banaszek_photon_2003,rehacek_multiple-photon_2003,the_multiple-photon_2007,webb_photostatistics_2009}. However, if the pulse energy remains above the saturation level, then the click probability of the initial bins will be close to unity. 
In general, we can calculate the click probability of a particular bin $j$, which must account for the coupling to previous detection bins and the number of photons in the incoming pulse. The detection probability of bin $j$ is
\begin{equation}\label{eqn:Pjn}
p_j=\left(1-\nu_j\right)\sum_{n=0}^\infty P\left(j|n\right)\rho_\textrm{in}\left(n\right)+\nu_j,
\end{equation}
where $P\left(j|n\right)$ is the probability that bin $j$ is occupied with at least one photon, given an incident number of photons $n$, $\rho_\textrm{in}\left(n\right)$ is the photon number probability distribution of the input state and $\nu_j\ll1$ is the dark count probability. 
A general expression for $P\left(j|n\right)$ for $j\geq 2$ is, in each case
\begin{equation}\label{eqn:pjn}
P\left(j|n\right)=
\begin{cases}
1-\left[1-\left(R\eta\right)^{j-1}\eta \left(1-R\right)\right]^n&\textrm{active}~,\\
1-\left[1-\left(R\eta\right)^{j-1}R^{-1}\left(1-R\right)^2\right]^n&\textrm{passive}~,
\end{cases}
\end{equation}
where $\eta$ is the loop roudtrip efficiency.
The probability for the first bin, $P\left(1|n\right)$, also depends on whether the loop is actively or passively switched. This may be written:
\begin{equation}
P\left(1|n\right)=\Big\{\begin{array}{ll}
1-\left[1-\eta \left(1-R\right)\right]^n&\textrm{active}~,\\
1-(1-R)^n&\textrm{passive}~.
\end{array}
\end{equation}

\begin{figure}[ht]
    \centering
    
		\subfloat[\label{fig:distsA}
	]{\includegraphics[width=0.27\linewidth]{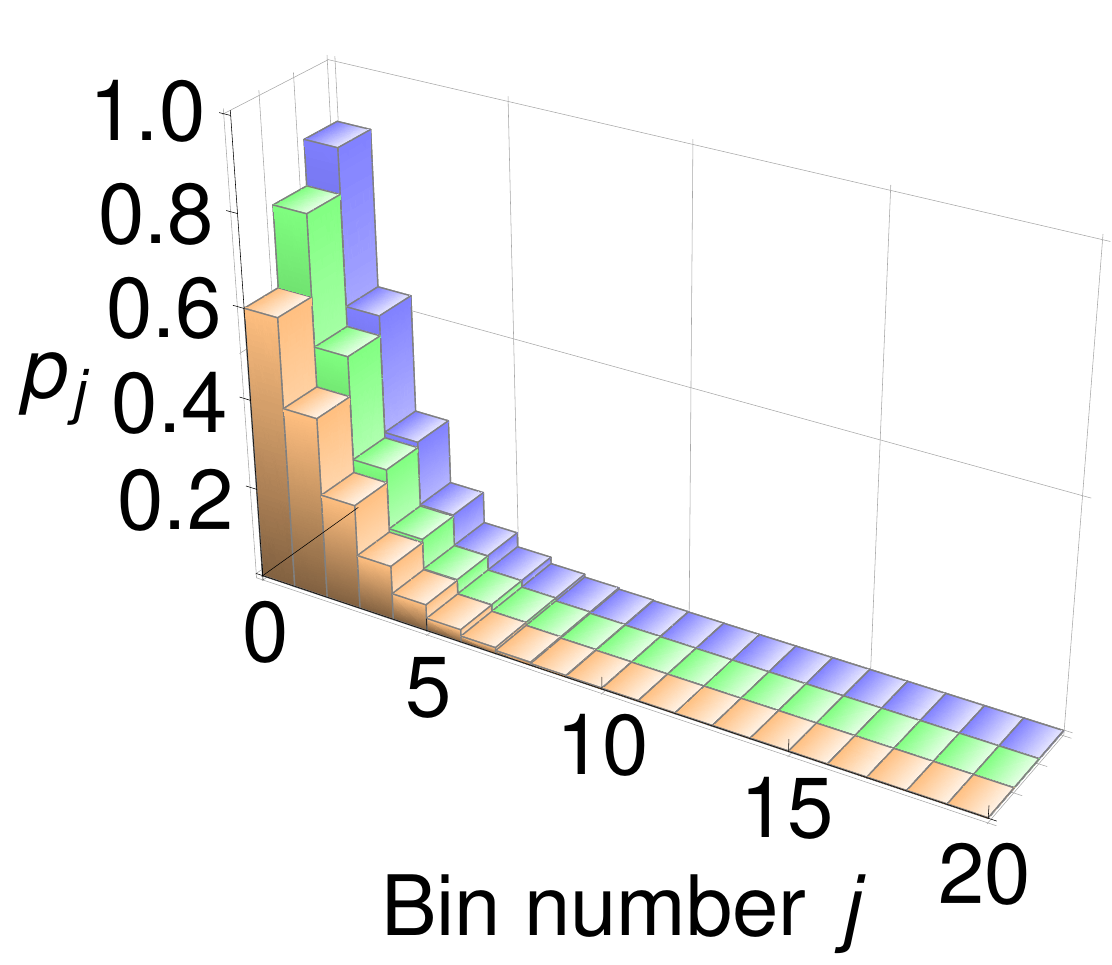}}\hspace{1.4cm}
		\subfloat[\label{fig:distsB}
		]{\includegraphics[width=0.27\linewidth]{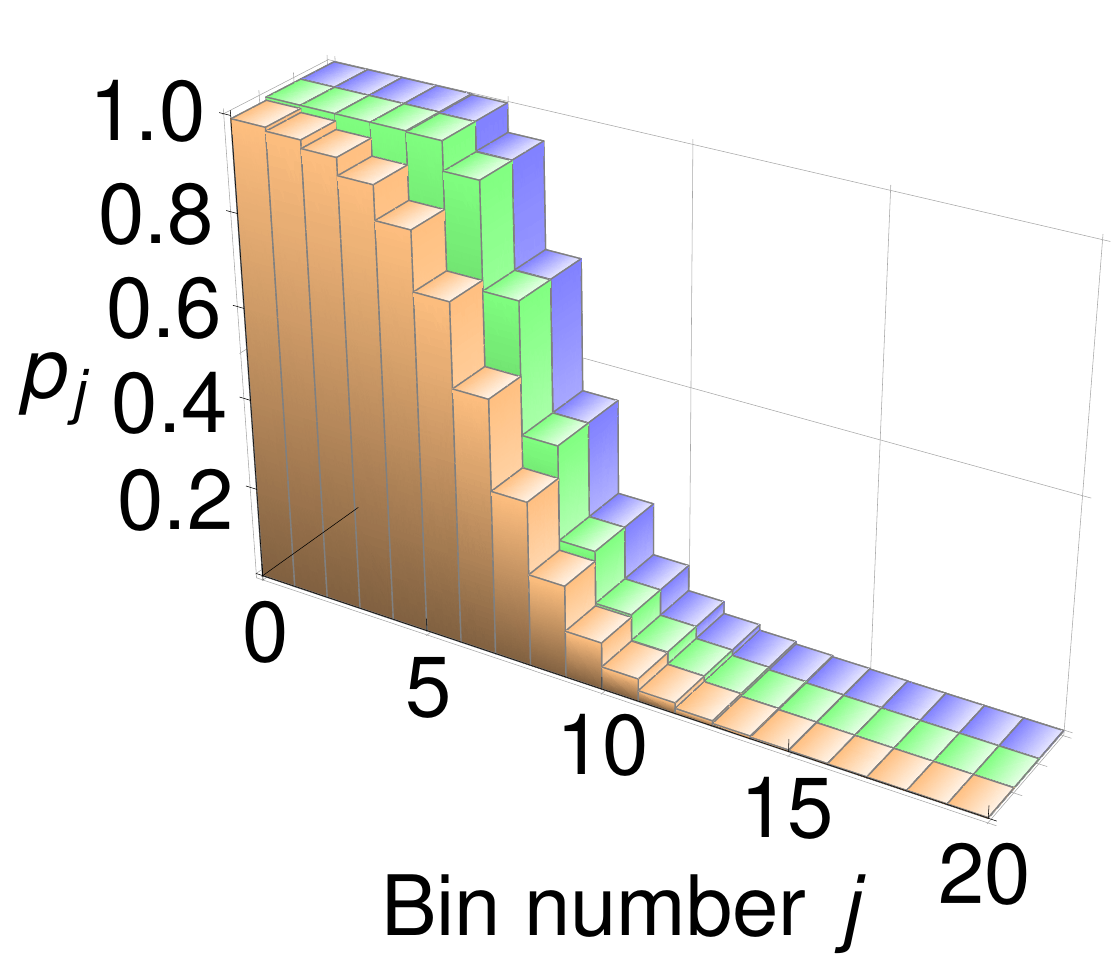} }
\caption{(a) Click probability $p_j$ for a passive loop detector as a function of bin $j$, for a Fock state (blue), coherent state (green) and thermal state (orange), each with a mean photon number of $\bar{n}=3$, outcoupling parameter $R=0.5$, loop efficiency $\eta=0.9$ and noise contribution $\nu=10^{-3}$. (b) As above, but with the mean photon number $\bar{n}=300$ in each case (all other parameters remain the same). In general, each of these classes produces slightly different distribution of clicks per bin $p_j$, for fixed mean photon number. This occurs due to the nonlinear response of the click detectors, and is most clearly identified close to the saturation regime ({e.g.} around bin 5) in (b).}\label{fig:distributions}
\end{figure}

From these expressions, $p_{j}$ can be analytically expressed in closed form for several classes of optical states, namely photon number (Fock) states, coherent states and thermal states, in terms of their mean photon number $\bar{n}$. Analytic forms for each state, in the case of active and passive loop architecture, are given in the Appendix. 
In the low mean photon number regime, all classes produce probabilities $p_{j\geq2}$ which are modelled by an exponential decay, an example of which is shown in Fig.~\ref{fig:distributions}(a). However, as the mean photon number of the incident light increases, saturation of the early bins occurs, leading to photon detection probability histograms such as in Fig.~\ref{fig:distributions}(b).

Note, however, that despite this saturation, the power incident on the detector can still be inferred directly from the data. 
This allows single-photon-level detectors to increase their effective sensitivity towards very bright states. 

\section{Bright-light sensitivity}
\begin{figure}[!h]
\centerline{\includegraphics[width=\linewidth]{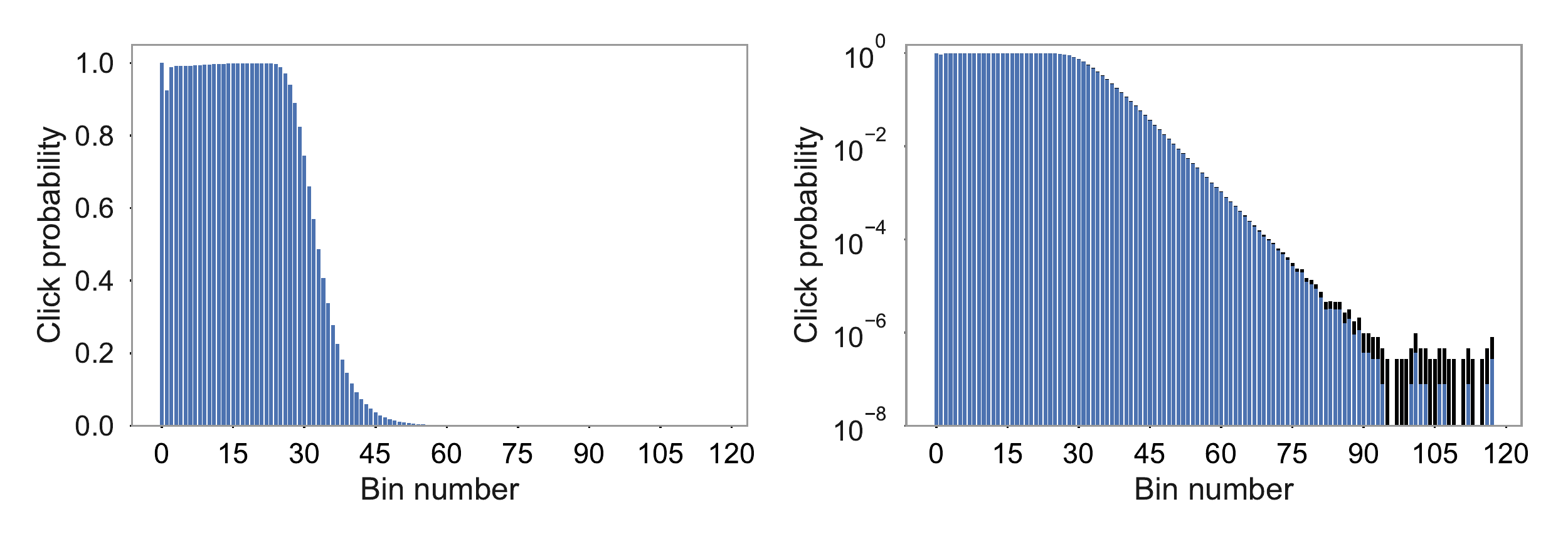}}
\vspace{-1.5em}
\subfloat[]{\includegraphics[width=0.5\linewidth]{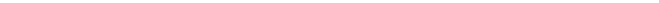}}
\subfloat[]{\includegraphics[width=0.5\linewidth]{w.png}}
\caption{Histogram of click probabilities on linear and log scales, for a coherent state input with an average of $\sim$250 000 photons per pulse to the passive loop detector with a repetition rate of \unit[50]{kHz}. Error bars are based on estimating the success probability of a Binomial distribution (shown in black).}
\label{fig:ringdown}
\end{figure}

To demonstrate sensitivity to bright light, we conducted an experiment in which a pulsed laser with a repetition rate of \unit[50]{kHz} was connected to a passive loop structure (for more details see Appendix). After the loop, \unit[1.61$\pm$0.08]{nW} average power  was measured with a power meter, corresponding to 251000$\pm$12500 photons per pulse. 
An example of the resulting histogram of the bin click probabilities $p_j$ is shown in Fig.~\ref{fig:ringdown}. 
From this histogram, and Eq.~(\ref{eqn:pjn}), we can determine the mean number of photons per pulse measured by the click detector to be $\bar{n}_\textrm{out}=208011\pm460$ with a relative error of $\sigma_{\bar{n}}/\bar{n}$ of 0.22\%. The ratio between $\bar{n}_\textrm{out}$ and the input mean photon number, as measured by the power meter, gives the detection efficiency of $82.8\%$. This procedure does not require cascaded calibrated attenuators~\cite{marsili_detecting_2013}; further details on the calibration procedure and error analysis are provided in the Appendix. Moreover, as an alternative to a bright light standard, one could use this device to bridge to the bright light regime using a known single-photon detector calibrated to a quantum luminescence standard, {i.e.} a correlated single photon source arising from parametric down-conversion~\cite{klyshko_use_1980,migdall_absolute_1995}.

The dynamic range of the detector is defined by $DR=10\log_{10}\left(\bar{n}/\nu\right)=10\log_{10}\left(n_\textrm{max}/\nu\eta R\right)$, where $n_\textrm{max}$ is determined by the damage threshold of the detector used, {i.e.} the pulse energy that causes the detector to become unresponsive to subsequent pulses. Using the bright measurement above, with an input photon number per pulse of $\bar{n}=2.5\times10^{5}$ and a the lower limit (minimum sensitivity) given by the noise floor ($\nu=1.2\times10^{-7}$ photons per pulse), we therefore demonstrate a dynamic range of \unit[123]{dB}.

\section{Measuring nonclassical signatures}
In addition to measuring bright states of light, the detector is also sensitive to nonclassical features of the input photon distribution; correlations between different click events in different bins can yield further insight. There exists extensive analysis of statistics arising from multiplexed detectors which can determine whether or not the underlying photon statistics are consistent with a classical distribution~\cite{haderka_direct_2005,webb_photostatistics_2009,kiesel_complete_2012,
sperling_sub-binomial_2012,sperling_true_2012,bartley_direct_2013,harder_time-multiplexed_2014,liu_photon-number_2014,chrapkiewicz_photon_2014,
sperling_uncovering_2015,lee_sub-poisson-binomial_2016,sperling_identification_2017,
bohmann_incomplete_2018}. One important example is the sub-binomial parameter~\cite{sperling_sub-binomial_2012}, which translates sub-Poissonian statistics of light into sub-Binomial click statistics. However these analyses typically rely on equal splitting between bins. To overcome this limitation, the generalisation by Lee and co-workers~\cite{lee_sub-poisson-binomial_2016} leads to the definition of the Poisson-binomial parameter
\begin{equation}\label{eqn:QPB}
Q_{PB}=N\frac{\langle\left(\Delta c\right)^2\rangle}{\langle c\rangle\left(N-\langle c\rangle\right)-N^2\sigma^2}-1~,
\end{equation}
where $c_k$ describes the probability of $k$ bins firing, which has a mean, and variance, given by
\begin{equation}
\langle c\rangle=\sum_{k=0}^Nkc_k;~\langle\left(\Delta c\right)^2\rangle=\sum_{k=0}^N\left(k-\langle c\rangle\right)^2c_k
\end{equation}
and
\begin{equation}
m=\frac{1}{N}\sum_{j=1}^Np_j;~\sigma^2=\frac{1}{N}\sum_{j=1}^N\left(p_j-m\right)^2~.
\end{equation}
This describes the mean probability and variance, respectively, of a bin providing a click event, given individual bin click probabilities $p_j$, for a total of $N$ bins. For uniform splitting among the bins, $m=p_j$ and $\sigma^2=0$, such that Eq.~(\ref{eqn:QPB}) reduces to the binomial parameter $Q_B$~\cite{sperling_sub-binomial_2012}. A sufficient condition for nonclassical light is negativity of the $Q_{PB}$ parameter.

\begin{figure}[!h]
\centerline{\includegraphics[width=0.8\linewidth]{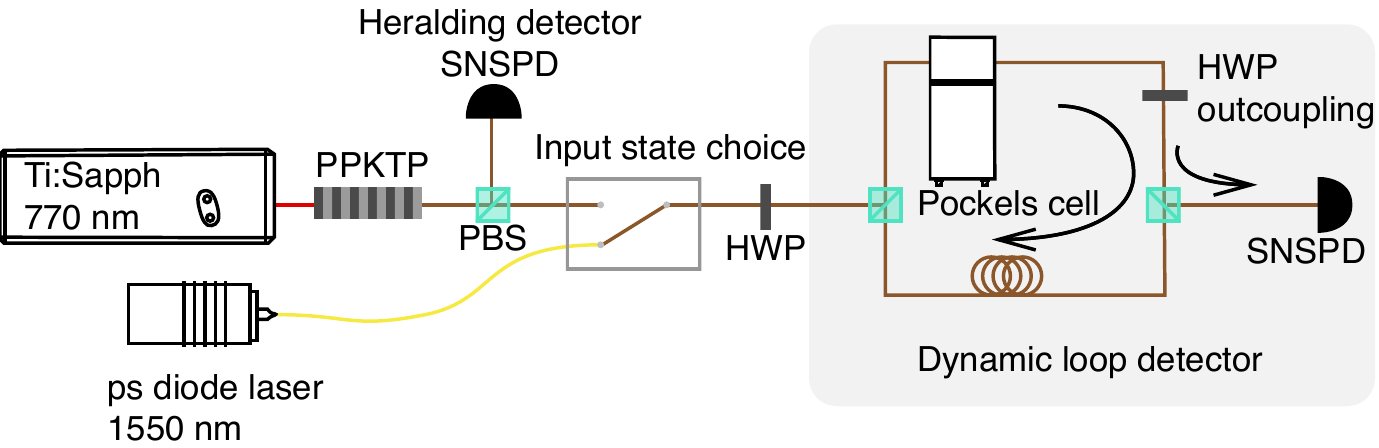}}
\caption{Detection of sub-binomial light with a dynamic loop detector. Heralded single photons and thermal states (from Periodically-poled Potassium Titanyl Phosphate - PPKTP waveguide), and coherent states are coupled into the loop with a fast Pockels cell polarization switch. Then on each pass through the loop a small fraction of the light, determined by the angle of the half-wave plate (HWP), is coupled out to the superconducting nanowire single photon detector (SNSPD).}
\label{fig.setup}
\end{figure}
\begin{figure}[!h]
\centering
\subfloat[\label{fig:QPBactive}]{\includegraphics[width=0.5\linewidth]{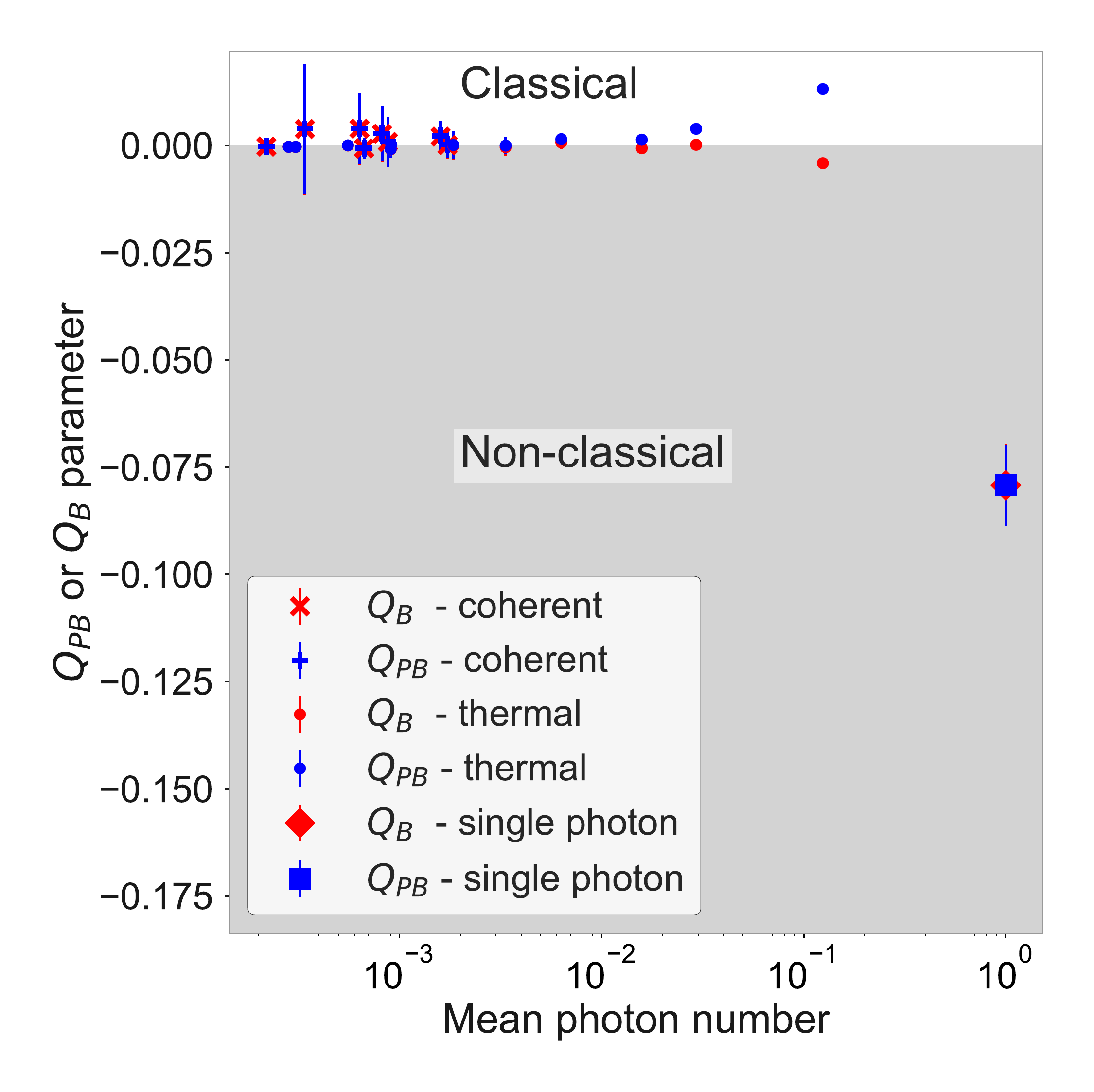}}
\subfloat[\label{fig:QPBpassive}]{\includegraphics[width=0.5\linewidth]{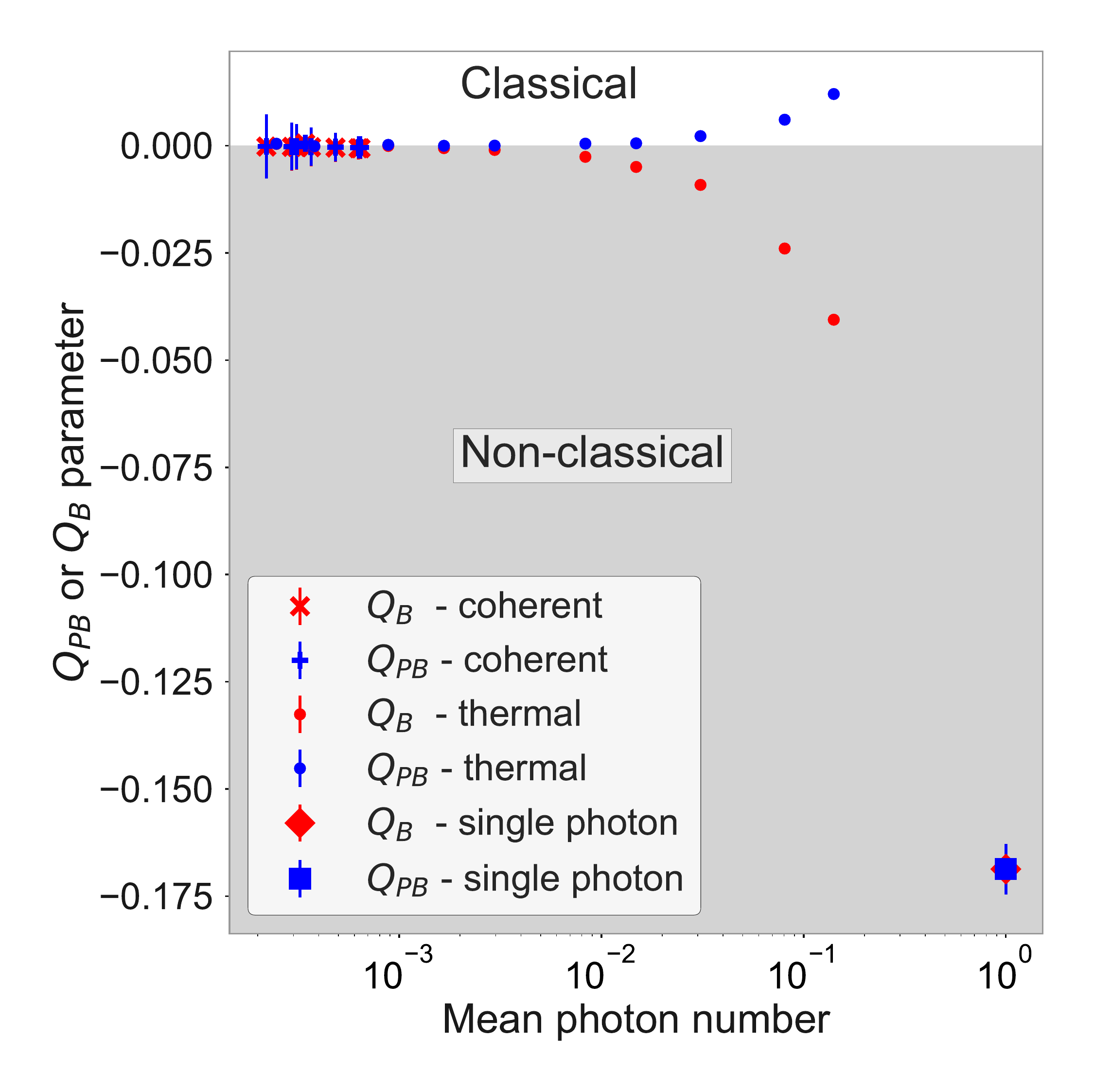}}
\caption{Classicality parameters $Q_{PB}$ and $Q_{B}$ as measured by the loop detector for heralded PDC states, for (a) actively switched loop (with coupling parameter $R=0.1$); and (b) passively switched loop (with $R=0.5$). Error bars are based on $10000$ Monte-Carlo-Simulations where the mean and variance are given by the measured $c_k$ distributions.}\label{fig.qpb}
\end{figure}

To demonstrate its ability to distinguish nonclassical light, we performed an experiment (Fig. \ref{fig.setup}) with three different types of states over a broad range of power levels power levels input to both active and passive implementations of the loop detector.  
We present the $Q_{PB}$ and $Q_B$ parameters as a function of mean photon number for heralded single photons, thermal states and coherent states in an actively-switched Fig.~\ref{fig.qpb}(a) and passively-switched Fig.~\ref{fig.qpb}(b) loop in. 
The multi-thermal states present statistics somewhere between thermal and Poissonian statistics, due to more than a single spectral mode emerging from the crystal. We estimate from our pump bandwidth and crystal parameters a spectral purity of $57\%$, or a Schmidt number of $K=1.8$ contributing thermal modes. The classicality of the multi-thermal photon statistics, in combination with the exponentially decaying $p_j$ probabilities, is not captured by a naive application of the $Q_B$ parameter to our data, as it shows negativity (nonclassicality). The $Q_{PB}$ parameter, on the other hand, handles this effect as expected. Both parameters correctly identify the nonclassicality of the heralded signal photons. 

\section{Conclusion}
In summary, we have demonstrated how multiplexing a single binary detector can be used to measure light levels across a large dynamic range (123 dB)
. We have used this device to perform precise calibration, as well as identify nonclassical phenomena in the click statistics. This high dynamic range can be used in a broad range of applications, for example precision absorption spectroscopy and characterisation of high-extinction components across a broad range of energies.

\section*{Appendix}
\setcounter{section}{0}
\def\thesection{\Alph{section}}
\section{Methods}

\textbf{Light sources}
For coherent states, we use a picosecond-pulsed semiconductor laser at \unit[1550]{nm} with a variable repetition rate, varying the mean photon number by attenuation. For multi-thermal states and heralded single photons, we pump a periodically-poled potassium titanyl phosphate waveguide designed for type II parametric down-conversion. The signal and idler modes (around \unit[1540]{nm}) are split on a polarization beam splitter (PBS), and the idler is detected to herald the signal single photons, or undetected for (nearly) thermal states in the signal mode. 

\textbf{Loop architecture}
For quantum characterisation, these states are sent to the active loop detector, which is made up of a free-space and fibre loop, with optional deterministic incoupling accomplished with an electro-optic Pockels cell~\cite{nitsche_quantum_2016}. The input light pulses are vertically polarized, then switched to horizontal with the Pockels cell to remain in the loop, passing through a fibre delay line (loop length \unit[480]{m}, corresponding to a delay of \unit[2.4]{$\upmu$s}). The outcoupling half-wave plate (HWP) sets the fraction of the light outcoupled on each round trip, by rotating some of the light to vertical polarization. Note that the outcoupling optics is not the same as the incoupling optic, resulting in slightly different loss for the first bin, compared to subsequent bins. The outcoupled light is sent to a commercial superconducting nanowire single photon detector (SNSPD) from Quantum Opus. The Klyshko efficiency of the photon (when using down-conversion) coupled directly through the loop (half round-trip) is $(20\pm2)\%$, and the round trip loop transmission is $(82\pm2)\%$.
Depending on whether the Pockels cell was activated, we implement either the active or passive loop detector structure.

For the high dynamic range measurements, we implemented a passive loop architecture constructed exclusively from optical fibres (loop length \unit[30]{m}, corresponding to a delay of \unit[156]{ns} between bins) and a low-loss fibre beam splitter. All components were polarization-maintaining, such that the (polarization dependent) detector efficiency is constant for each bin, and the beam-splitter ratio (which is also typically polarization-dependent) is constant for each round trip. The round trip loop transmission in this case was 86\%, which includes additional (artificially-induced) loss to produce a faster decay rate and therefore allows for a higher repetition rate. In these experiments, we used a commercial superconducting nanowire single photon detector (SNSPD) from Photon Spot.

\textbf{Detector and data acquisition}
In both the active and passive experiments, the detector output was recorded on a time-tagger, allowing for all events to be recorded. Gating windows of \unit[4]{ns} were applied in post processing to reduce the effect of dark counts. The gating windows are separated by the corresponding loop delay, corresponding to the time for the pulse to propagate the length of the fibre loop, and well above the dead time of the detector. 

\section{Analytic expressions for bin click probabilities}
Based on Eq.~(1) and expressions for the photon number distributions $\rho_\textrm{in}$ for Fock states $\rho_\textrm{Fock}=\delta_{\bar{n},n}$, coherent states $\rho_\textrm{coh}=\exp\left[-\bar{n}\right]\tfrac{\bar{n}^n}{n!}$ and thermal states $\rho_\textrm{therm}=\frac{\bar{n}^n}{\left(1+\bar{n}\right)^{n+1}}$, and whether the loop is actively or passively switched, the following analytic expressions for the bin click probability $p_j$ can be derived:

\textit{Active:}
\begin{align}
p_{j}^\textrm{Fock}&=
1-\left(1-\nu\right)\left[1-\left(1-R\right)R^{-1}\left(R\eta\right)^j\right]^{\bar{n}}\\
p_{j}^\textrm{coh}&=
1-\left(1-\nu\right)\exp\left[-\left(1- R\right)R^{-1}\left(\eta R\right)^j\bar{n}\right]\label{eqn:pjcohxac}\\
p_{j}^\textrm{therm}&=
1-\left(1-\nu\right)\frac{R}{R+\left(1-R\right)\left(R\eta\right)^j\bar{n}}
\end{align}

\textit{Passive:}
\begin{align}
p_{j}^\textrm{Fock}&=
\begin{cases}
1-\left(1-\nu\right)(1-R)^{\bar{n}}&j=1\\
1-\left(1-\nu\right)\left[1-\left(1-R\right)^2R^{-1}\left(R\eta\right)^{j-1}\right]^{\bar{n}}&j\geq2
\end{cases}
\\\label{eqn:PassCoh}
p_{j}^\textrm{coh}&=
\begin{cases}
1-\left(1-\nu\right)\exp\left[-R\bar{n}\right]&j=1\\
1-\left(1-\nu\right)\exp\left[-\left(1-R\right)^2R^{-1}\left(\eta R\right)^{j-1}\bar{n}\right]&j\geq2
\end{cases}
\\
p_{j}^\textrm{therm}&=
\begin{cases}
1-\left(1-\nu\right)\frac{1}{1+R\bar{n}}&j=1\\
1-\left(1-\nu\right)\frac{R^2\eta}{R^2\eta+\left(1-R\right)^2\left(R\eta\right)^j\bar{n}}&j\geq2
\end{cases}
\end{align}

In all cases we assume a mean photon number $\bar{n}$ and a constant dark-count probability per bin (noise floor) $\nu_j=\nu$ for all time bins. 

\section{Calibration procedure}
\begin{figure}[!h]
\centerline{\includegraphics[width=.8\linewidth]{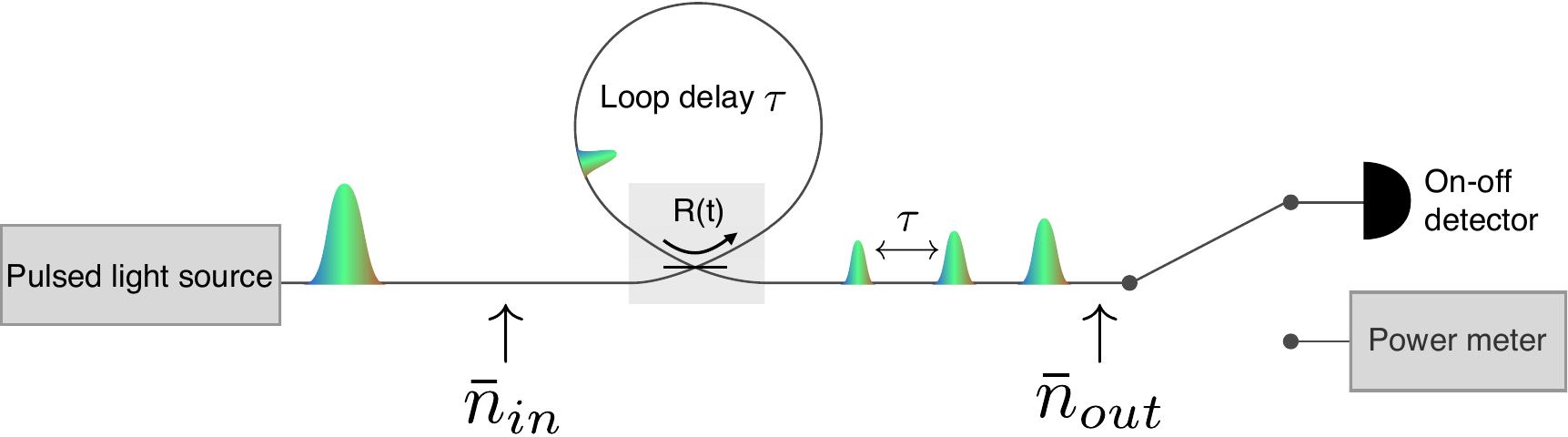}}
\caption{Schematic of calibration setup. $\bar{n}_\textrm{in}$ indicates the mean photon number per pulse incident on the loop. $\bar{n}_\textrm{out}$ is the mean photon number after the loop, summed over all bins. $\bar{n}_\textrm{out}$ can be switched between an on-off detector or a power meter. Reliable switching is important for the calibration procedure as systematic errors can be produced during this process.} 
\label{fig:calibration_overview}
\end{figure}

The high dynamic range of the detector can be used to bridge the gap between bright light and quantum-based optical power standards. To calibrate the detector efficiency to a known bright light standard, we suggest the following procedure, using the setup shown in Fig.~\ref{fig:calibration_overview}. 
Given a coherent state with a mean photon number determined with a power meter $\bar{n}_\text{PM}=\textrm{power}/(\textrm{photon energy}\times 
\textrm{repetition rate})$, one can define a system detection efficiency $\eta_\textrm{SDE}$ as
\begin{equation}\label{eqn:SDE}
\eta_\textrm{SDE}=\frac{\bar{n}_\textrm{measured}-n_\textrm{dark counts}}{\bar{n}_\textrm{PM}}
\end{equation}
where $\bar{n}_\textrm{measured}$ is calculated from on-off detector response at the output of the loop detector.

To calculate $\bar{n}_\textrm{measured}$, we begin with the mean photon number per bin $\bar{n}_\text{j}$. For the active case this is given by 
\begin{equation}
\bar{n}_\text{j} = \bar{n}_\text{in} (1-R)R^{j-1} \eta^j~,
\end{equation}
and for the passive case 
\begin{align}
\bar{n}_\text{j}=\Big\{\begin{array}{ll}
\bar{n}_\text{in} R&j=1\\
\bar{n}_\text{in}(1-R)^2 R^{j-2}\eta^{j-1}&j\geq2
\end{array}
\end{align}
where, in both cases $\bar{n}$ is the number of photons per pulse present prior to entering the loop.

With these equations we can also calculate the total outgoing mean photon number
\begin{equation}
\bar{n}_\text{out}=\sum_{j=1}^\infty \bar{n}_\text{j}.
\end{equation}
Which means for the active case 
\begin{equation}
\label{eq:pow_in_out_active}
\bar{n}_\text{out}= \frac{\bar{n}_\text{in}(-1+R) \eta }{-1+R \eta}
\end{equation}
and for the passive case 
\begin{equation}
\label{eq:pow_in_out_passive}
\bar{n}_\text{out}= \frac{\bar{n}_\text{in}(-R - \eta + 2 R \eta) }{-1+R \eta}.
\end{equation}

\begin{figure}[ht!]
\subfloat[]{
\includegraphics[width=.5\linewidth]{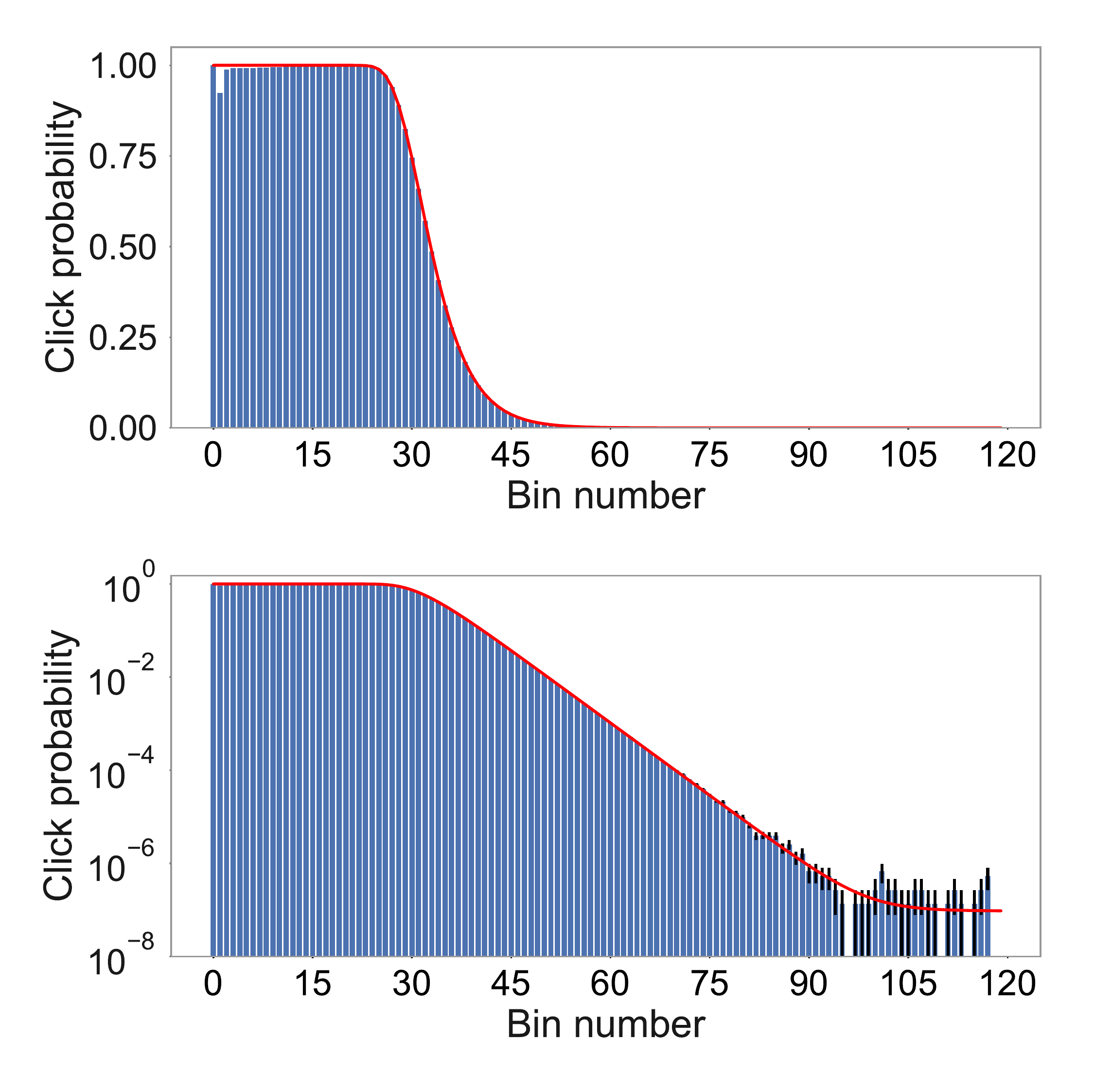}}
\subfloat[]{
\includegraphics[width=.5\linewidth]{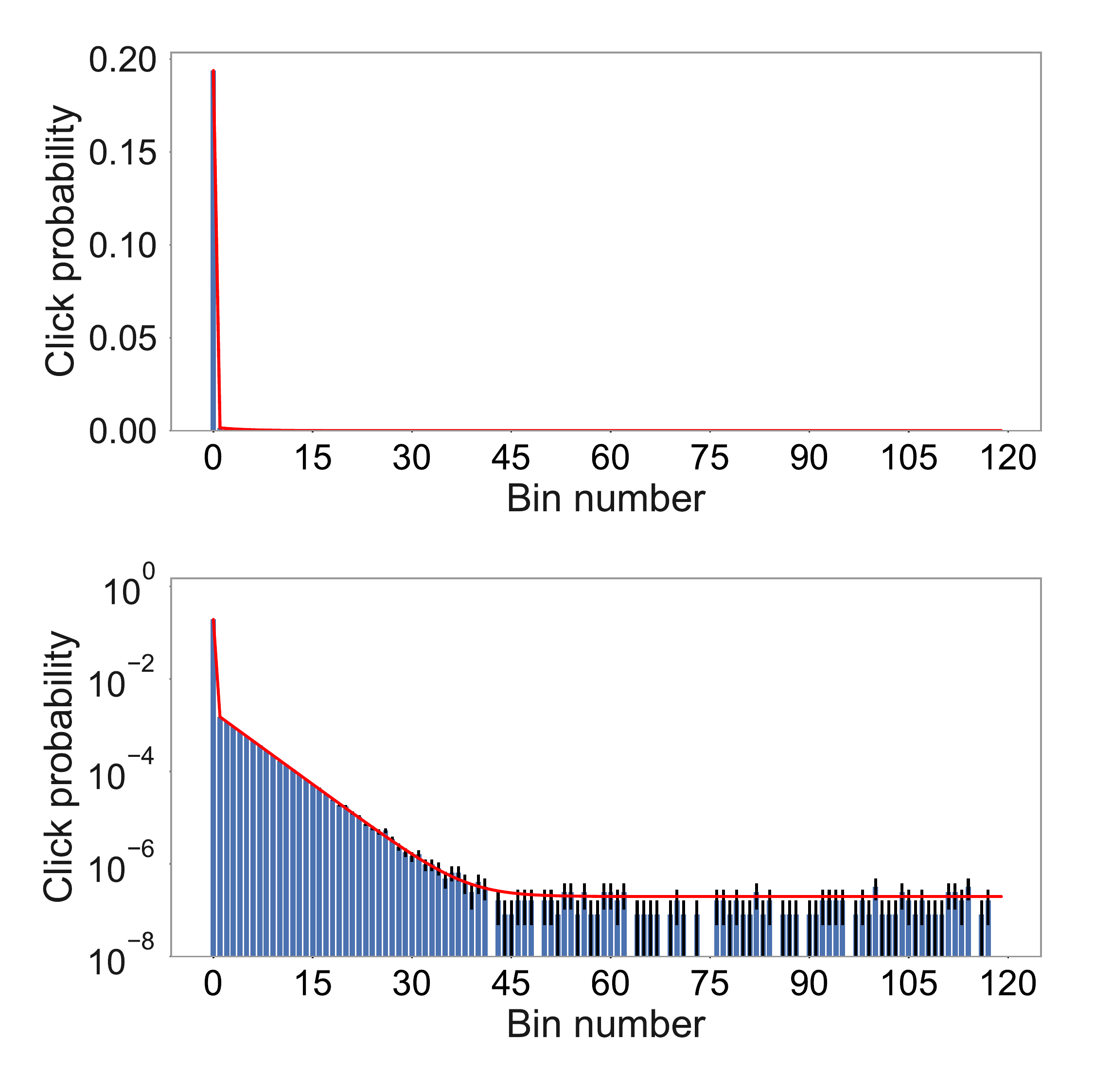}}

\caption{Histogram showing the bin probability in linear (top) and logarithmic (bottom) scale for a calibration measurement (a) and a highly attenuated input (b), using a passive loop. Red curves: Fit based on Eq.~(\ref{eqn:PassCoh}). The fit for the attenuated coherent state gives an estimate for $R = 0.91370 \pm 5\cdot 10^{-5}$ and $\eta = 0.8615\pm 3\cdot 10^{-4} $.}
\label{fig:calibration}
\end{figure}

\begin{figure}[!h]
\centerline{\includegraphics[width=.9\linewidth]{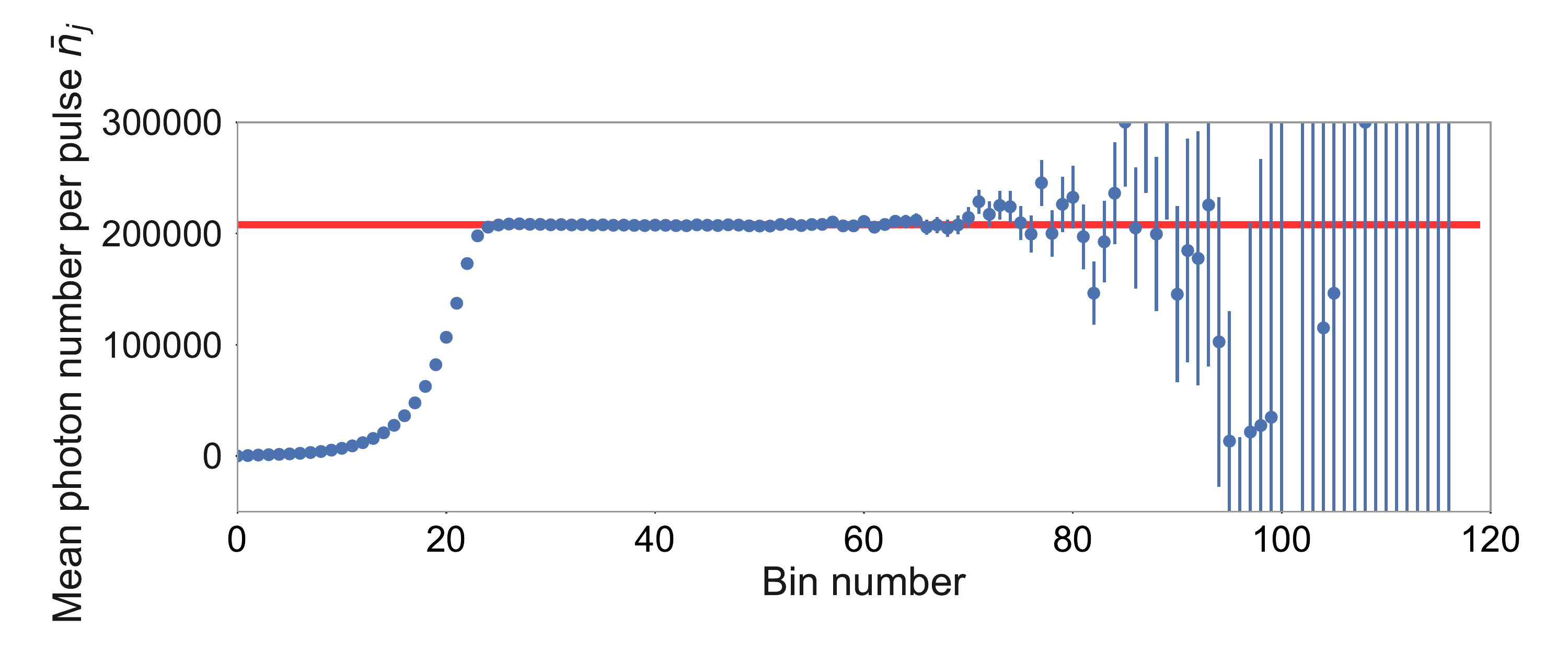}}
\caption{Inferred mean photon number per pulse $\bar{n}_\text{j}$ after the loop from measured click probabilities $p_\text{j}$. Red line: mean value of $\bar{n}_\textrm{measured}=208011$. The first bins $j\leq25$ are excluded because dead times from back reflections reduced the count rate (see next section for further details).} 
\label{fig:nfrompj}
\end{figure}

Experimentally we can now measure the click probability per bin $p_\text{j}$ with the click detector under test (cf. Fig.~\ref{fig:calibration}(a)). Using Eqs.~(\ref{eqn:pjcohxac}) and (\ref{eqn:PassCoh}) (and substituting $\bar{n}=\bar{n}_\textrm{in}$) for coherent states as well as Eqs.~(\ref{eq:pow_in_out_active}) and (\ref{eq:pow_in_out_passive}) each bin probability $p_\text{j}$ gives an estimate for the mean photon number per pulse $\bar{n}_\text{out}\vert_{j}$ (see Fig.~\ref{fig:nfrompj}). 
For the active case we can write 
\begin{equation}
\label{eq:n_out_active}
\bar{n}_\text{out}\vert_{j}= \frac{R \eta  (R \eta )^{-j} \ln\left[\frac{1-\nu}{1-p_j}\right]}{1-R \eta}
\end{equation}
and for the passive case 
\begin{equation}
\label{eq:n_out_passive}
\bar{n}_\text{out}\vert_{j}= 
\begin{cases}
\frac{( R +\eta-2 \eta R) \ln\left[\frac{1-\nu}{1-p_\text{j}}\right]}{R (1-R \eta )}&j=1\\
\frac{ (R+\eta-2\eta R)(R\eta)^{1-j} R \ln\left[\frac{1-\nu}{1-p_\text{j}}\right]}{(1-R \eta )(R-1)^2 }&j\geq2~.
\end{cases}
\end{equation}

To arrive at a final estimate of the mean photon number $\bar{n}_\textrm{measured}$, we use the arithmetic mean weighted by the experimental error in the individual estimates of $\bar{n}_\textrm{out}|_j$, as determined above. This provides our unbiased estimator of $\bar{n}_\textrm{out}$, with the weights given by the inverse of the variances  $\sigma_{\bar{n}_\textrm{out}|_j}^2$ (see next section for details on how these errors are calculated) of each measurement of $\bar{n}_\textrm{out}|_j$, {i.e.}
\begin{equation}
\bar{n}_\textrm{measured}=\frac{\sum_{j}w_j\bar{n}_\textrm{out}|_j}{\sum_{j}w_j}~,
\end{equation}
where
\begin{equation}
w_j=\frac{1}{\sigma_{\bar{n}_\textrm{out}|_j}^2}~.
\end{equation}
Fig.~\ref{fig:nfrompj} shows the measured values for $\bar{n}_\text{out}\vert_{j}$ and $\bar{n}_\textrm{measured}$. The limits on the summation over $j$ are chosen such that the errors in the individual measurements of $\bar{n}_\textrm{out}|_j$ are not underestimated ({i.e.} that their uncertainties are reliable). From Fig.~\ref{fig:ExpError}, this is from bin 25 onwards. The use of the weighted arithmetic mean penalises values with high error bars; therefore including high bin numbers, where the uncertainty is high, does not limit our precision.

With this summation ($j\geq25$), the mean photon number per pulse is $\bar{n}_\text{measured}=208011\pm460$. The value of $\bar{n}_\text{measured}$  can now be compared to the value from the power meter $n_\textrm{PM}=251000\pm12500$ to calculate the System Detection Efficiency (SDE), {i.e.} Eq.~(\ref{eqn:SDE}). The results in an $\eta_\textrm{SDE}=82.8\pm4\%$. The dominant error comes from the power meter reading, not $\bar{n}_\text{measured}$, and we neglect errors in laser drift and fibre coupling efficiency. The value of $\eta_\textrm{SDE}$ includes the fibre into the cryostat and the detector efficiency. It can be seen from Eq.~(\ref{eq:n_out_active}) that the value of $\bar{n}_\text{out}$ for the active case only depends of the product $\eta R$. This product is well known as the slope of the histogram in the decay region (logarithmic plot) is $1-\eta R$. The passive case, however, is different. Here the individual values of $\eta$ and $R$ must be known to infer $n_\text{out}$.  Therefore we need an additional measurement step for the passive case in order to determine at least one of $R$ and $\eta$. 
We suggest to add an unknown attenuator before or after the loop such that the first bin is non-saturated. An example measurement can be seen in Fig.~\ref{fig:calibration}(b). By fitting this histogram with Eq.~(\ref{eqn:PassCoh}) values for $R$ and $\eta$ can be determined. The full information about these two values is contained in the jump from the first bin to the second as well as in following linear decay region. Therefore the attenuator must be strong enough that the first bin in non-saturated (otherwise the uncertainties in the click probability are too high, {c.f.} next section) and low enough to see the linear decay region before the dark count level. The advantage of this method is that the beam splitter reflectivity $R$ and the loop efficiency $\eta$ can be determined without changing the loop. 

It should be noted that losses before the loop and after the loop (up to the point where either a power meter or a click detector is placed) do not matter for the calibration procedure as this is a rescaling of the power that is already considered by the power meter reading.

This method can be easily altered to measure the power of bright pulses if the SDE is known, for example to bridge a quantum standard to bright light.
\section{Estimation of uncertainty}
\begin{figure}[!h]
\centerline{\includegraphics[width=.9\linewidth]{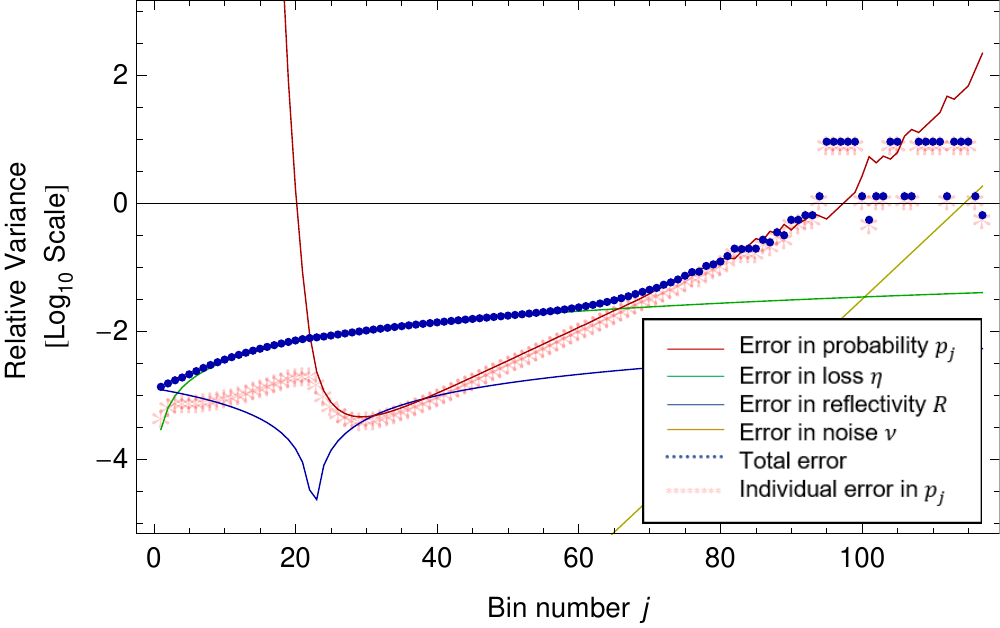}}
\caption{Relative uncertainty contributions $\sqrt{\frac{\sigma^2_i}{\bar{n}^{\prime2}}\left(\frac{\d \bar{n}^\prime}{\d i}\right)^2}$ from each error source, arising from uncertainty in the measured bin probabilities $p_j$ (red line), the loop loss $\eta$ (green line), the loop reflectivity $R$ (blue line) and the noise $\nu$ (gold line). Note that these quantities are evaluated using the a posteriori weighted mean $\bar{n}^\prime$ (i.e.) the mean photon number averaged over all $j $. The blue dots are the total relative error contributions, and red crosses the relative error in $p_j$, evaluated at the individually determined photon number $\bar{n}_\textrm{out}|_j$ (in contrast to the solid lines). The deviation of these measures at low bin number $j\lesssim25$ arises from undercounting caused by back-reflection-induced inefficiency, resulting in an underestimate of the error associated with $p_j$. This is discussed further in the following section.}
\label{fig:ExpError}
\end{figure}

The relative uncertainty in $\bar{n}_\textrm{measured}$ arises from the uncertainty in the component quantities from which $\bar{n}_\textrm{out}$ is calculated. For each bin $j$, we obtain an independent estimate of $\bar{n}_\textrm{out}$, {i.e.} $\bar{n}_\textrm{out}\rvert_j$. This is given by Eq.~(\ref{eq:n_out_passive})
The relative uncertainty in each of these estimates of $\bar{n}_\textrm{out}|_j$ is given by Gaussian error propagation:
\begin{align}\nonumber
\frac{\sigma_{\bar{n}_\textrm{out}\rvert_j}}{\bar{n}_\textrm{out}\rvert_j}=&\left.\left[\frac{\sigma_{p_j}^2}{\bar{n}_\textrm{out}^2\rvert_j}\left({\frac{\d \bar{n}_\textrm{out}}{\d p_j}}\right)^2+\frac{\sigma_{R}^2}{\bar{n}_\textrm{out}^2\rvert_j}\left(\frac{\d \bar{n}_\textrm{out}}{\d R}\right)^2+\frac{\sigma_{\eta}^2}{\bar{n}_\textrm{out}^2\rvert_j}\left(\frac{\d \bar{n}_\textrm{out}}{\d \eta}\right)^2+\frac{\sigma_{\nu}^2}{\bar{n}_\textrm{out}^2\rvert_j}\left(\frac{\d \bar{n}_\textrm{out}}{\d \nu}\right)^2\right]^{1/2}\right\rvert_j\\\nonumber
=&\Bigg\{\sigma_{p_j}^2\frac{1}{\left(1-p_j\right)^2\ln\left(\frac{1-\nu}{1-p_j}\right)^2}\\\nonumber
&+\sigma_{R}^2\left[\frac{1-j}{R}+\frac{1}{R\left(1-R\eta\right)}+\frac{1}{R\left(1-2R\right)+\eta\left(1-2R\right)^2}-\frac{2R}{1-3R+2R^2}\right]^2\\\nonumber
&+\sigma_{\eta}^2\left[\frac{1-j}{\eta}+\frac{\left(1-R\right)^2}{\left(R+\eta-2R\eta\right)\left(1-R\eta\right)}\right]^2\\\label{eqn:errors}
&+\sigma_{\nu}^2\frac{1}{\left(1-\nu\right)^2\ln\left(\frac{1-\nu}{1-p_j}\right)^2}\left.\Bigg\}^{1/2}\right\rvert_j~.
\end{align}
The absolute errors are all experimentally determined: $\sigma_R,~\sigma_\eta$ are determined by the fitting procedure, explained in the previous section. $\sigma_\nu$ is determined from Poissonian error from the dark-count probability measurements. Only $\sigma_{p_j}$, determined by Poisson error in the click statistics, depends on the bin number; the other absolute errors are constant for each bin. For the data shown above, the absolute and relative uncertainties are given in table \ref{table1}:
\begin{table}[htbp]
\caption{\bf Absolute and relative errors of the measured values}
\centering
\begin{tabular}{cccc}
Quantity&Value&Absolute error $\sigma_i$&Relative error $\sqrt{\frac{\sigma_i}{\bar{i}}}$\\
\hline
$R$&$0.91370$&$5\times10^{-5}$&$0.006\%$\\
$\eta$&$0.8615$&$3\times10^{-4}$&$0.034\%$\\
$\nu$&$1.20\times10^{-7}$&$2\times10^{-9}$&$1.9\%$\\
\hline
\end{tabular}
\label{table1}
\end{table}
The contribution of each source of error to the relative error in the photon count rate to be determined is shown graphically in Fig.~\ref{fig:ExpError}. The blue dots correspond to evaluating Eq.~(\ref{eqn:errors}) 
 as a function of the bin number $j$, to show the overall uncertainty in determining $\bar{n}_\textrm{out}|_j$ at each bin. The solid lines correspond to evaluating the individual contributions $\sqrt{\frac{\sigma^2_i}{\bar{n}_\textrm{out}^2}\left(\frac{\d \bar{n}_\textrm{out}}{\d i}\right)^2}$ for $i\in\left\{p_j,R,\eta,\nu\right\}$. Note that we use the weighted arithmetic mean as our estimate of the mean photon number $\bar{n}_\textrm{out}$, rather than the individually determined estimates for $\bar{n}_\textrm{out}\rvert_j$, since these estimates under-count at low bin numbers, for reasons described in the section on limits and assumptions, below. As a comparison, we show with red crosses $\sqrt{\frac{\sigma^2_{p_j}}{\bar{n}_\textrm{out}^2|_j}\left.\left(\frac{\d \bar{n}_\textrm{out}}{\d i}\right)^2\right\rvert_j}$, {i.e.} the relative uncertainty contribution due to $p_j$, evaluated at the individual bin number $j$, and not the weighted average. The result of the undercounting shows a clear underestimate of the relative uncertainty at $j\lesssim25$.

The resulting uncertainty of the weighted arithmetic mean is thus given by the inverse root of the sum of the weights, i.e.
\begin{equation}
\sigma_{\bar{n}^\prime}=\frac{1}{\left(\sum_jw_j\right)^{1/2}}~.
\end{equation}
Given the weights calculated above, and summing from bin 25, we obtain a relative error of $\sigma^2_{\bar{n}}/\bar{n}^{\prime 2}=0.22\%$ in our determination of the mean photon number after the loop.

The uncertainty in determining an unknown System Detection Efficiency (SDE) stems from uncertainties in its constituent quantities. We compare our method to that used in ~\cite{marsili_detecting_2013}, in which the expression for the SDE is $SDE=PCR/\left[P_Ca_2,a_3R_{\textrm{SW}}/\left(1-\rho\right)/E_\lambda\right]$, given by the photon count rate $PCR$, the power on the power meter $P_c$, the attenuations $a_2,a_3$, the switch ratio $R_\textrm{SW}$, the reflectivity of the fibre connected to the switch $\rho$, and the photon energy $E_\lambda$. The principle uncertainties arise in the quantities $PCR$, $P_C$, $a_{2,3}$ and $R_\textrm{SW}$. The relative uncertainties in $P_C$ and $R_\textrm{SW}$ remains present in both experiments. To evaluate each approach, we compare the relative error contributions arising from calibrated attenuation and determination of the photon count rate. In~\cite{marsili_detecting_2013} (using the values in Table SI 2 and Eq.~(1)), this contribution is the root quadrature sum of $\sigma_\textrm{PCR}/PCR=0.14\%$ and $\sigma_\alpha/\bar{n}=0.20\%$. Since two attenuators were used, this value contributes twice to the overall relative uncertainty. Thus the relative uncertainty arising from these two sources is $\sqrt{\left(\sigma_\textrm{PCR}/PCR\right)^2+2\left(\sigma_\alpha/\alpha\right)^2}=0.32\%$. Our method does not use calibrated attenuators, thereby removing this source of measurement error, to comparable error therefore arises in $\sigma_{\bar{n}}/\bar{n}=0.22\%$, as indicated above. Our method therefore provides a significant reduction in the error when compared with using calibrated attenuators.

\begin{figure}[!h]
\centerline{\includegraphics[width=.7\linewidth]{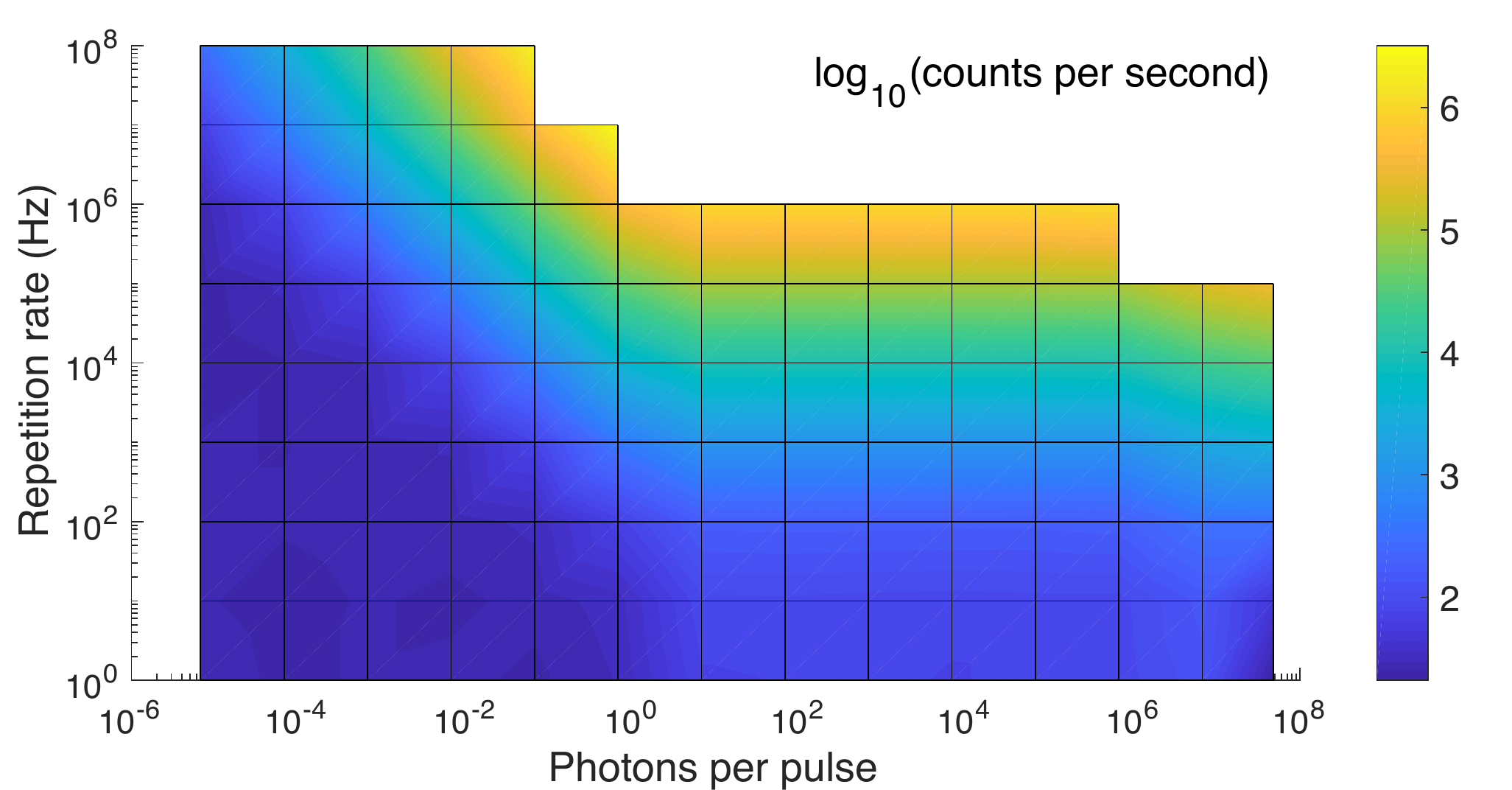}}
\caption{Single bin click probability as a function of incident number of photons per pulse and pulse repetition rate. Once latching occurs, no counts can be measured (upper right corner).}
\label{fig:latching}
\end{figure}

The precision with which the inference of input mean photon number detected via the loop can be achieved depends on both the fitting error for determining $R$ and $\eta$, as well as the statistical error from the number of data points used. Both of these depend on the number of occupied bins above the noise floor of the device. For noise $\nu\ll 1$, we can approximate the number of occupied bins above the noise floor as
\begin{equation}\label{eqn:maxbins}
j_\textrm{max}\approx\frac{\log_{10}\left[\frac{\nu}{n_\textrm{max}}\right]}{\log_{10}\left[\eta\left(1-R\right)\right]}~.
\end{equation}
The upper limit to the dynamic range $n_\textrm{max}$ is determined by the damage threshold of the detector used, {i.e.} the pulse energy that causes the detector to become unresponsive to subsequent pulses. For SNSPDs the first failure mode is latching at high count rates, where the device creates a self-sustaining hotspot and cannot reset without manual reduction of the bias current. We have found a large region of pulse energy (1 to $10^7$ photons per pulse) where the latching point depends only on the repetition rate and not on the pulse energy (see Fig.~\ref{fig:latching}). Thus SNSPDs are ideal candidates for the loop detector, as they can handle many photons in the first bins and still be single-photon sensitive with low noise in the later bins. We have tested up to $\bar{n}=4.8\times10^{7}$ photons per \unit[100]{ps} pulse at \unit[100]{kHz} repetition rate with a WSi SNSPD, and found no latching or degradation of performance.

\section{Limitations and assumptions}
\begin{figure}[!h]
		\subfloat[\label{fig:BackRefTTs}
	]{\includegraphics[width=.5\linewidth]{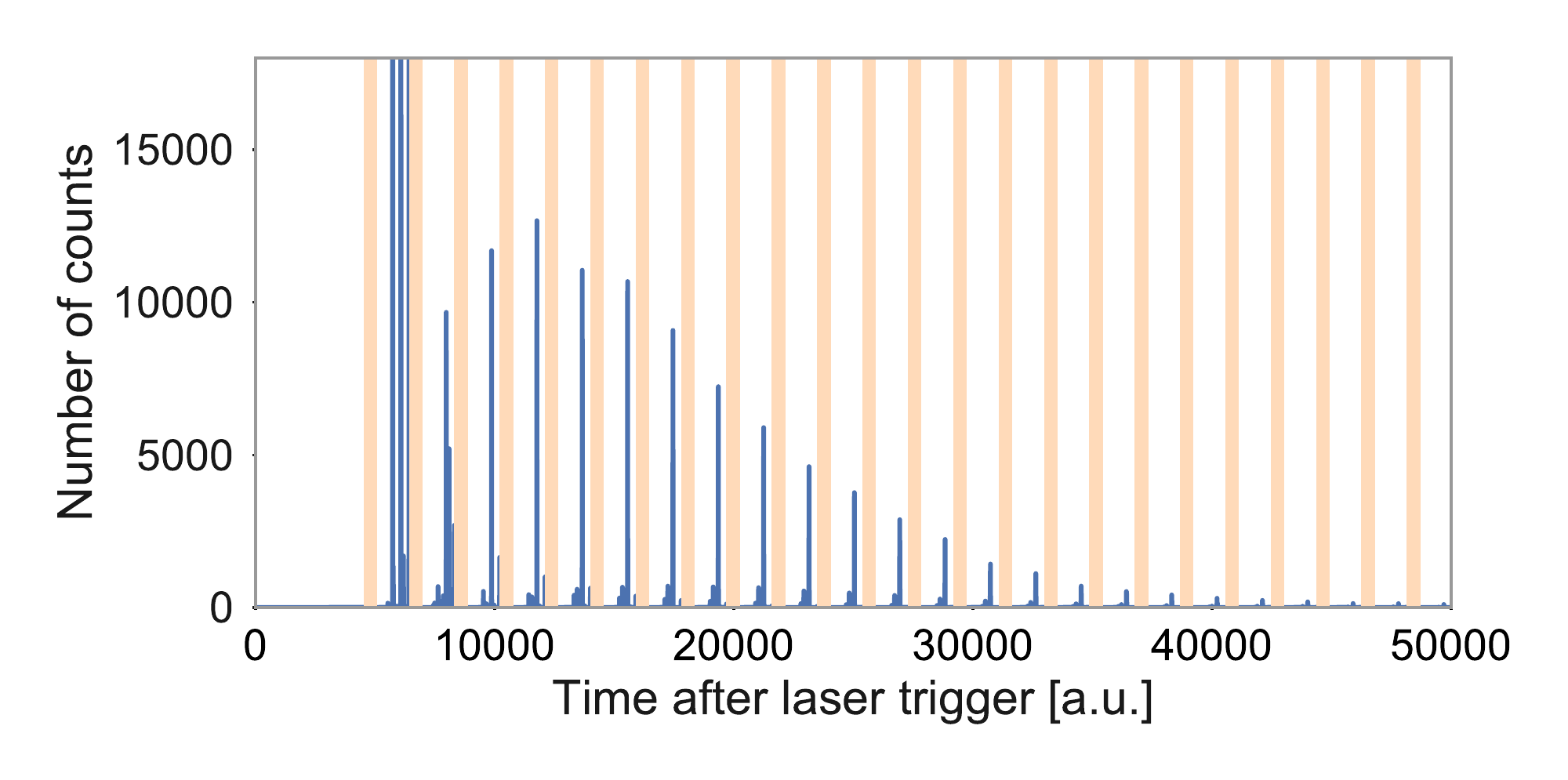}}
		\subfloat[\label{fig:BackRefHist}
		]{\includegraphics[width=.5\linewidth]{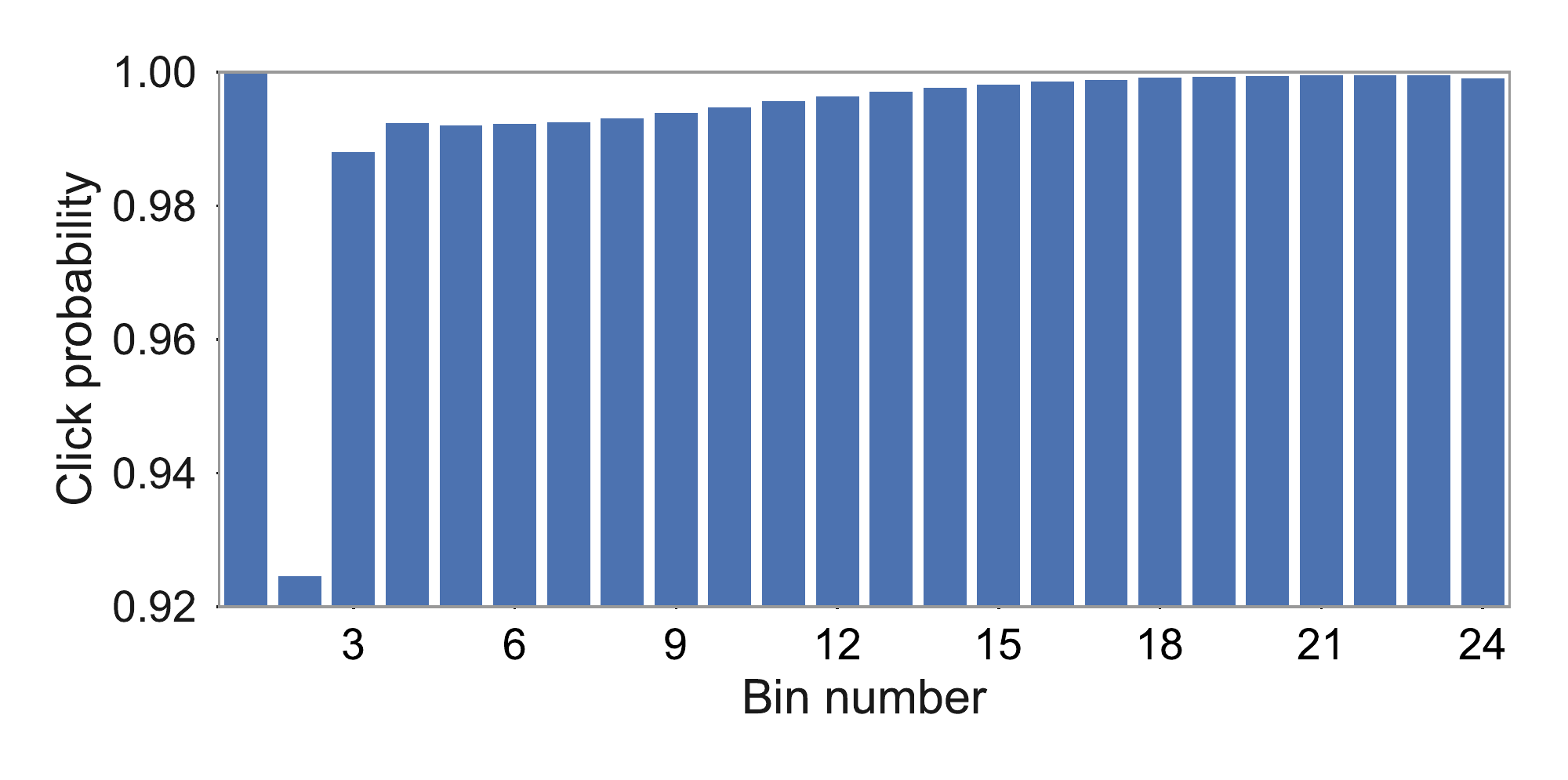}}
\caption{(a) Raw time-tags for a bright pulse, zoomed in on the first 24 bins, showing clear excess counts outside of the expected loop bins (orange). This arises from spurious back-reflections present in the setup. (b) Click probability as a function of bin number. In early bins $2\leq j\lesssim20$, back reflections cause a reduction in the expected counts since they induce dead time, which can be seen by a deviation from the expected click probabilities.}\label{fig:BackRefs}
\end{figure}

The procedure we present relies on a few important assumptions to be applied correctly. Most generally, the detector efficiency to be calibrated must remain constant for all bins. Effects which change this assumption should be avoided, or where possible, appropriately accounted for. We neglect random drift of the efficiency or laser power, since the timescale over which we take data is relatively short (ca. \unit[150]{s}). More critical is bin-to-bin variations in efficiency. This could be caused by a different polarization for each bin, and an associated polarization-dependent efficiency. We obviate this error source by using polarization-maintaining fibre components to build our loop. A further source of error arises from missed counts due to back-reflection-induced dead time losses. In early bins, there are a great many photons circulating in the setup. While most photons are in the loop, a small fraction may be reflected back towards the detector, out of synch with the actual photons to be measured. Whilst these will be detected outside the acceptance window, their detection will cause some dead time, during which time the detector is ``blind'' to further photons. This effect is visible in the raw time-tag data, shown in Fig.~\ref{fig:BackRefs}(a). 
This plot is effectively an optical-time-domain-reflectometery (OTDR) spectrum, evaluated over many orders of magnitude. It shows counts appearing outside the expected bins (shown in orange).

Thus, true events may be missed. This is effect is clearly seen in Fig.~\ref{fig:BackRefs}(b), where the second bin in particular is significantly lower than expected. Reducing back reflections is therefore crucial to the reliable operation of this system.

\section{Trade-off between operating speed and dynamic range}\label{sec:limits}
In the case of SNSPDs, the upper limit on the dynamic range is determined by the number of occupied bins per unit time that the detector can handle before it latches. Crucially, over a certain range of repetition rates, it does not depend on the power delivered the detector. We demonstrate this by varying the repetition rate and mean photon number per pulse incident on the detector. The resulting count rate is shown in Fig.~\ref{fig:latching}.

One might expect that the product of photons per pulse $\bar{n}$ and repetition rate $r$ ({i.e.} $\bar{n}\times r=P$, the power delivered to the detector) would be constant, with a corresponding constant latching threshold. This is true up to a mean photon number of one photon per pulse, where the latching threshold is $\sim10^6$ photons per second or \unit[0.1]{pW} at a wavelength of \unit[1550]{nm}. However, above this, the mean photon number per pulse plays almost no part in determining the latching threshold. This allows us to put much more incident power on the detector before it latches, provide the repetition rate is slightly lower. The highest we achieved is $5\times10^7$ photons per pulse at a repetition rate of \unit[100]{kHz}, yielding an optical power of \unit[0.6]{$\upmu$W}. This measurement was limited by available laser pulse energy; higher power may well be achievable. The consequences of this effect may also be interesting in terms of the detector dynamics and heat dissipation models, which clearly behave nonlinearly with incident optical power.

\section*{Funding}\par
European Union Horizon 2020 665148 (QCUMbER); DFG (Deutsche Forschungsgemeinschaft) SFB/TRR 142; Natural Sciences and Engineering Research Council of Canada;  European Commission ERC project QuPoPCoRN (725366). \\


\end{document}